\documentstyle[12pt]{article}
\begin{document}
\thispagestyle{empty}
\begin{flushright}
INR 0988/98 \\
September 1998 \\
\end{flushright}
\bigskip\bigskip\bigskip
\vskip 2.5truecm
\begin{center}
{\LARGE {Ginsparg-Wilson relation and}}
\vskip 7pt
{\LARGE {lattice Weyl fermions}}
\end{center}  
\vskip 1.0truecm
\centerline{S. V. Zenkin}
\vskip5mm
\centerline{Institute for Nuclear Research of
the Russian Academy of Sciences}
\centerline{60th October Anniversary Prospect 7a, 117312 Moscow,
Russia}
\centerline{E-mail: zenkin@al20.inr.troitsk.ru, zenkin@ms2.inr.ac.ru}
\vskip 2cm
\bigskip \nopagebreak \begin{abstract}
\noindent
We demonstrate that in the topologically trivial gauge sector the
Ginsparg-Wilson relation for lattice Dirac operators admits an
exactly gauge invariant path integral formulation of the Weyl
fermions on a lattice.
\end{abstract}
\vskip 1.5cm

\newpage\setcounter{page}1

{\bf 1.} The Ginsparg-Wilson relation \cite{GW} for chirally
noninvariant lattice Dirac operator yields perfect solution to the
long-standing problem of maintaining chiral symmetry in lattice
vector gauge theories \cite{Ha1}--\cite{Lu1}. 
In turn, the chiral noninvariance of the lattice Dirac operator is 
necessary for the operator to be local on a regular lattice without 
producing species doubling \cite{NiN}, and also to reproduce the 
Atiyah-Singer index theorem \cite{it} on a finite lattice \cite{Ze}.
The simplest form of the relation is
\begin{equation}
D \gamma_5 + \gamma_5 D = 2 r D \gamma_5 D,
\label{GW}
\end{equation}
where $D$ is the lattice Dirac operator and $r$ is a nonzero real
parameter (the lattice spacing $a$ is set to one). Then, if $D$
satisfies (\ref{GW}), the fermion action
\begin{equation}
S_f = \overline{\psi} D \psi,
\label{sf}
\end{equation}
is invariant under chiral transformations of the following form 
\begin{eqnarray}
&& \psi \rightarrow \exp [i \alpha + i \beta \gamma_5 E(s)] \psi, \cr
&& \overline\psi \rightarrow \overline\psi \exp [ -i \alpha 
+ i \beta F(s) \gamma_5 ],
\label{sym}
\end{eqnarray}
where $\alpha$ and $\beta$ are the parameters of the vector and the 
axial transformations and  
\begin{eqnarray}
&& E(s) = 1 - (2r-s) D, \cr
&& F(s) = 1 - s D,
\label{gen}
\end{eqnarray}
is a family of the generators of the axial transformations
parametrised by the real parameter $s$. The symmetry (\ref{sym}),
(\ref{gen}) at $s = r$, in which case $E = F = 1-rD$, has been
discovered by L\"uscher \cite{Lu1}, and at $s = 0$ considered in
\cite{Ni}, \cite{KY}. Obviously, it protects the theory from the
additive mass renormalization.  At the same time the measure in the
fermion path integral
\begin{equation}
Z_f = \int \prod_{x} d\psi(x) d\overline{\psi}(x) \exp(-S_f)
= \mbox{det} D,
\label{Z}
\end{equation}
is not invariant under (\ref{sym}), and the corresponding Jacobian
produces the anomaly in the divergence of the axial part of the
N\"other current associated with these transformation \cite{Lu1}. The
anomaly is proportional to \cite{HLN}, \cite{Lu1}, \cite{Lu2}
\begin{equation}
q(x) = i \frac{1}{2} \frac{\partial}{\partial \beta(x)} 
J^{-1}(\beta) = -r \mbox{tr} (\gamma_5 D).
\end{equation}

Another remarkable property of solutions of the Ginsparg-Wilson
relation is that \cite{HLN}, \cite{Lu1}
\begin{equation}
\sum_x q(x) = n_+ - n_- \equiv - \mbox{index }D,
\label{ind}
\end{equation} 
where $n_{+}$ ($n_{-}$) is the number of zero modes of the $D$ of
positive (negative) chirality. So the Atiyah-Singer index theorem
\cite{it}
\begin{equation}
\mbox{index }D = Q,
\label{it}
\end{equation}
where $Q$ is topological number of the background gauge field, 
may be reproduced exactly even on a finite lattice.

At present only one explicit solution of the Ginsparg-Wilson equation
is known, the operator proposed by Neuberger \cite{Ne}, that indeed
reproduces the index theorem (\ref{it}) \cite{Ne}, \cite{Ch} and is
local at least in sufficiently smooth gauge field background
\cite{HJL}.

Since $[\gamma_5 E(0)]^2=[F(0) \gamma_5]^2 =1$, one can defined two
pairs of projecting operators \cite{Ni}
\begin{eqnarray}
&&P^{E}_{\pm} = \frac{1}{2}[1 \pm \gamma_5 E(0)] = \frac{1}{2}[1 \pm 
\gamma_5 (1-2rD)], \cr
&&P_{\pm} = \frac{1}{2}[1 \pm F(0) \gamma_5] = \frac{1}{2}(1 \pm
\gamma_5),
\end{eqnarray}
and the action (\ref{sf}) can be written as \cite{Ni}
\begin{equation}
S_f = \overline{\psi}_+ D \psi^{E}_{+} + \overline{\psi}_- D
\psi^{E}_{-},
\label{sfc}
\end{equation}
where
\begin{equation}
\psi^{E}_{\pm} = P^{E}_{\pm} \psi, \quad
\overline{\psi}_{\pm} = \overline{\psi} P_{\mp}.
\label{lr}
\end{equation}
The decomposition (\ref{sfc}), (\ref{lr}) itself however does not yet
allow one to write down the functional integral (\ref{Z}) as the
product of two factors related to the Weyl fermions of opposite
chiralities \cite{Na}, since the fermion measure in (\ref{Z}) is not
factorised according to (\ref{sfc}), and the fields $\psi^{E}_{\pm}$
cannot be treated as the Weyl fields.

We are aiming now to demonstrate that at one condition specified
below, a certain change of variables $\psi$ leads to natural definition
of the path integral for the Weyl field of positive or negative
chirality. \\

{\bf 2.} To proceed, let us set some properties of the solutions of
relation (\ref{GW}). We limit our consideration to the operators $D$
with the property
\begin{equation}
D^{\dagger} = \gamma_5 D \gamma_5.
\label{hc}
\end{equation}
In the chiral representation of $\gamma$ matrices where $\gamma_5 = 
\mbox{diag}(1, -1)$,
$\gamma_{\mu}^{\dagger} = \gamma_{\mu}$, such operators have the form
\begin{equation}
D = \pmatrix {M & D_{-} \cr D_{+} & M \cr}.
\label{D}
\end{equation}
Matrices $D_{+}$ and $D_{-} = -D^{\dagger}_{+}$ are lattice
transcriptions of the covariant Weyl operators $\sigma_{\mu}
(\partial_{\mu} + i A_{\mu})$ and $\sigma_{\mu}^{\dagger}
(\partial_{\mu} + i A_{\mu})$, respectively, where $A_{\mu}$ is the
gauge field, $\sigma_{\mu} = (1, i)$ in two dimensions and
$\sigma_{\mu} = (1, i\sigma_i)$ in four dimensions, and the matrix
$M$ determines chirally noninvariant part of $D$. In this
representation
\begin{equation}
\psi = \pmatrix {\chi_+ \cr \chi_- \cr}, \quad \overline{\psi} 
= (\chi^{\dagger}_{-} \,\,\, \chi^{\dagger}_{+}),
\label{chi}
\end{equation}   
where $\chi_{\pm}$ are the Weyl fields of the chirality $\pm$.

Then from the Ginsparg-Wilson relation (\ref{GW}) it follows
\begin{eqnarray}
&&D_+ D_- = D_- D_+ = M(M - r^{-1}), \cr
&&MD_{\pm} = D_{\pm}M,
\label{pro}
\end{eqnarray}
i.e. all the entries of matrix
(\ref{D}) commute with each other. Using these properties it is easy
to obtain the following relations
\begin{eqnarray}
&&\mbox{det}D = \mbox{det}(r^{-1} M), \\  
\label{det}
&&\mbox{det}(1-rD) = \mbox{det}(1-rM).
\label{nec}
\end{eqnarray}
Note that singularity of the matrix $1-rD$, and therefore of the
matrix $1-rM$, is the necessary condition \cite{CZ} for the operator
$D$ to have nonzero index (\ref{it}).

Consider the combination $1-rM$ in more detail. Using the fact
\cite{Na}, \cite{CZ} that any solution of eqs.\ (\ref{GW}),
(\ref{hc}) can be presented in the form
\begin{equation}
D=\frac{1}{2r}(1+V), \quad V^{\dagger} = \gamma_5 V \gamma_5,
\end{equation}
where $V$ is a unitary matrix, and taking into account (\ref{GW}) and
(\ref{pro}), we find that
\begin{equation}
\pmatrix {1-rM & 0 \cr 0 & 1-rM \cr} = \frac{1}{2} \left(1-
\frac{V+V^{\dagger}}{2} \right),
\label{nne}
\end{equation}
from where it follows that the matrix $1-rM$ is nonnegative.
Combining now this fact with eq.\ (\ref{it}) and the above mentioned
necessary condition \cite{CZ}, we conclude that the matrix $1-rM$ is
positive definite in the topologically trivial gauge sector $Q = 0$,
except may be some exceptional configurations on which it gets
singular.

Let us now consider the theory in the topologically trivial sector
with the exceptional configurations excluded. Then $1-rM$ is positive
definite and there exist the unique positive matrix $(1-rM)^{1/2}$
and the unitary matrix
\begin{equation}
U = \pmatrix {(1-rM)^{1/2} & -rD_{-}(1-rM)^{-1/2}  \cr 
-rD_{+}(1-rM)^{-1/2} & (1-rM)^{1/2}  \cr}, \quad U^{\dagger}U = 1, 
\quad \mbox{det}U = 1,
\label{U}
\end{equation}
such that the operator $D U^{\dagger}$ is chirally invariant: 
\begin{equation}
D U^{\dagger} = \pmatrix { 0 & D_{-}(1-rM)^{-1/2} \cr 
D_{+}(1-rM)^{-1/2} & 0 \cr}, \quad D U^{\dagger} \gamma_5 = 
\gamma_5 D U^{\dagger}.
\label{DW}
\end{equation}
Note that the $U$ diagonalizes the projecting operators 
$P^{E}_{\pm}$:  
\begin{equation}
U P^{E}_{\pm} U^{\dagger} = P_{\pm},
\label{dia}
\end{equation}
and that such a diagonalization is possible only in the sector 
$Q = 0$.

Now it is straightforward to redefine the variables $\psi$. Making
the obvious change of variables
\begin{equation}
\psi' \equiv \pmatrix {\chi_+' \cr \chi_-' \cr} = U \psi,
\end{equation}
whose Jacobian in view of (\ref{U}) equals unity, and omitting the
primes we get
\begin{eqnarray}
&&Z_f = Z_+ Z_-, \cr
&&Z_{\pm} = \int \prod_{x} d\chi_{\pm}(x) d\chi^{\dagger}_{\pm}(x)
\exp(-\chi^{\dagger}_{\pm} W_{\pm} \chi_{\pm}),
\label{Zpm}
\end{eqnarray}
where the Weyl operators $W_{\pm}$ read as
\begin{equation}
W_{\pm} = \frac{D_{\pm}}{\sqrt{1-rM}}.
\label{Wpm}
\end{equation}
This is an explicit path integral realization of the factorization of
$Z_f$ pointed out in \cite{Na}. Note that in view of (\ref{pro}) and
(\ref{det}) we do have $\mbox{det}W_+ \mbox{det}W_- = \mbox{det}D$,
and in the free fermion case the operators $W_{\pm}$ have correct
continuum limit.

Thus eqs.\ (\ref{Zpm}), (\ref{Wpm}) define the path integrals for the
Weyl fermions in the topologically trivial gauge sector $Q = 0$. We
should make however few important comments. \\

{\bf 3.} The most striking feature of this formulation is its exact 
gauge invariance. Indeed, the fermion measure and the actions in 
(\ref{Zpm}), (\ref{Wpm}) are invariant under the ordinary gauge 
transformations
\begin{equation}
\chi_{\pm}(x) \rightarrow g_{\pm}(x) \chi_{\pm}(x), \quad  
\chi^{\dagger}_{\pm}(x) \rightarrow \chi^{\dagger}_{\pm}(x) 
g^{\dagger}_{\pm}(x), \quad g^{\dagger}_{\pm}(x) g_{\pm}(x) = 1.
\label{gt}
\end{equation}
This means that in such a formulation the so called consistent
anomaly \cite{BZ} vanishes for each Weyl fermion. Also obviously,
that all the N\"other currents associated with transformations
(\ref{gt}) are conserved. We would like to emphasize however that
this does not contradict to the index theorem (\ref{ind}),
(\ref{it}), since the above formulation exists only in the
topologically trivial sector.

The operators $W_{\pm}$ in (\ref{Wpm}) are nonlocal. This follows
from the analysis of the Fourier transform of $W_{\pm}$,
$W_{\pm}(p)$, in the free fermion limit. Indeed, taking into account
the first equation in (\ref{pro}), one can see that $W_{\pm}(p)$ is
discontinuous at the boundary of the Brillouin zone and thus avoids
species doubling \cite{NiN}. An explicit form of the $W_{\pm}$ can be
constructed from Neuberger's operator \cite{Ne}. For instance, simple 
consideration of the chiral Schwinger model on a finite torus 
$L \times L$ in the constant gauge field background with such 
operators shows that $Z_{\pm}$ are real, and the exceptional 
configurations correspond to $eA_{\mu} = \pi/L$.

With one exception, all previous gauge invariant formulations with
nonlocal actions \cite{nl} failed due to non canceling singularities
in the interaction vertices \cite{nlf}. Due to such singularities
none of those formulations could reproduce correct absolute value of
the fermion determinant without special subtractions \cite{Sl} even
in the perturbation theory. The exception is the non-local fixed
point action \cite{Wi}, \cite{BW} corresponding to $r=0$ in
(\ref{GW}), whose explicit form however is known only in the lowest
orders of the perturbation theory. Thus, eqs.\ (\ref{Zpm}),
(\ref{Wpm}) present the first example of the nonlocal gauge invariant
formulation that succeeds to reproduce correct value of $|Z_{\pm}|$
nonperturbatively.

Of course, due to the limitation only to the topologically trivial
sector $Q = 0$, the formulation (\ref{Zpm}), (\ref{Wpm}) is not
complete. However within this sector the question of principle
arises: whether nonlocality of Weyl operators always produces some
defects, so far hidden for the $W_{\pm}$ in (\ref{Wpm}), or the
chiral anomalies in the topologically trivial sector is necessary
attribute of only local formulations. \\

I am grateful to T.~W.~Chiu for discussions on ref.\ \cite{Ni}.


\begin{thebibliography}{99}

\bibitem{GW} P.~H.~Ginsparg and K.~G.~Wilson, Phys. Rev. D25 (1982)
2649 
%
\bibitem{Ha1} P.~Hasenfratz, Nucl. Phys. B (Proc. Suppl.) 63A-C
(1998) 53 
%
\bibitem{HLN} P.~Hasenfratz, V.~Laliena and F.~Niedermayer, Phys.
Lett. B427 (1998) 125 
%
\bibitem{Ne} H.~Neuberger, Phys. Lett. B417 (1998) 141; ibid. B427
(1998) 353 
%
\bibitem{Ha2} P.~Hasenfratz, Lattice QCD without tuning, mixing and
current renormalization, hep-lat/9802007 
%
\bibitem{Lu1} M.~L\"uscher, Phys. Lett. B428 (1998) 312
%
\bibitem{NiN} H.~B.~Nielsen and M.~Ninomiya, Nucl. Phys. B185 (1981)
20 [E: B195 (1982) 541]; ibid. B193 (1981) 173 
%
\bibitem{it} M.~F.~Atiyah and I.~M.~Singer, Ann. Math. 87 (1968) 546;
ibid. 93 (1971) 139; 
A.~S.~Schwarz, Phys. Lett. 67B (1977) 172 
%
\bibitem{Ze} S.~V.~Zenkin, Phys. Rev. D58 (1998) 057505 
(hep-lat/9803002)
%
\bibitem{Ni} F.~Niedermayer, Exact chiral symmetry, topological
charge and related topics, plenary talk given at the International
Symposium on Lattice Field Theory, Boulder, July 13-18, 1998 
%
\bibitem{KY} Y.~Kikukawa and A.~Yamada, Axial vector current of exact
chiral symmetry on the lattice, hep-lat/9808026 
%
\bibitem{Lu2} M.~L\"uscher, Topology and the axial anomaly in abelian 
lattice gauge theories, hep-lat/9808021
%
\bibitem{Ch} T.~W.~Chiu, Topological charge and the spectrum of
exactly massless fermion on the lattice, hep-lat/9804016, to appear
in Phys. Rev. D 
%
\bibitem{HJL} P.~Hern\'andez, K.~Jansen and M.~L\"uscher, Locality 
properties of Neuberger's lattice Dirac operator, hep-lat/9808010
%
\bibitem{Na} R.~Narayanan, Ginsparg-Wilson relation and the overlap 
formula, hep-lat/9802018
%
\bibitem{CZ} T.~W.~Chiu and S.~V.~Zenkin, On solutions of the 
Ginsparg-Wilson relation, hep-lat/9806019
%
\bibitem{BZ} W.~A.~Bardeen and B.~Zumino, Nucl. Phys. B 244 (1984)
421 
%
\bibitem{nl} S.~D.~Drell, M.~Weinstein and S.~Yankielowicz, Phys.
Rev. D 14 (1976) 1627; 
C.~Rebbi, Phys. Lett. B 186 (1987) 200; 
S.~V.~Zenkin, Mod. Phys. Lett. A 6 (1991) 151 
%
\bibitem{nlf} L.~H.~Karsten and J.~Smit, Nucl. Phys. B 144 (1978)
536; Phys. Lett. B 85 (1979) 100; A.~Pelissetto, Ann. Phys. (N. Y.)
182 (1988) 177 %
\bibitem{Sl} A.~A.~Slavnov, Phys. Lett. B 348 (1995) 553; 
A.~A.~Slavnov and N.~V.~Zverev, Nonlocal lattice fermion models on 
the 2d torus, hep-lat/9710093, to appear in Theor. Math. Phys. 
%
\bibitem{Wi} U.-J.~Wiese, Phys. Lett. B 315 (1993) 417
%
\bibitem{BW} W.~Bietenholz and U.-J.~Wiese, Phys. Lett. B 378 (1996)
222

\end{thebibliography}
\end{document}